# Beyond the Dichotomy:
# How Ride-hailing Competes with and Complements Public Transport

Oded Cats[1*], Rafal Kucharski[1], Santosh Rao Danda[2] and Menno Yap[1]

[1] Department of Transport and Planning, Delft University of Technology, Delft, the Netherlands
[2] Uber BV, Amsterdam, Netherlands

*o.cats@tudelft.nl
Stevinweg 1, 2628 CN Delft, the Netherlands

ABSTRACT
Since ride-hailing has become an important travel alternative in many cities worldwide, a fervent debate is underway on whether it competes with or complements public transport services. We use Uber trip data in six cities in the United States and Europe to identify the most attractive public transport alternative for each ride. We then address the following questions: (i) How does ride-hailing travel time and cost compare to the fastest public transport alternative? (ii) What proportion of ride-hailing trips that do not have a viable public transport alternative? (iii) How does ride-hailing change overall service accessibility? (iv) What is the relation between demand share and relative competition between the two alternatives?

Our findings suggest that the dichotomy - competing with or complementing - is false. Though the vast majority of ride-hailing trips have a viable public transport alternative, between 20% and 40% of them have no viable public transport alternative. The increased service accessibility attributed to the inclusion of ride-hailing is greater in our US cities than in their European counterparts. Demand split is directly related to the relative competitiveness of travel times i.e. when public transport travel times are competitive ride-hailing demand share is low and vice-versa.

**Keywords:** Ride-hailing; Transportation Network Companies; Public Transport; Accessibility.

Introduction

Since their inception around a decade ago, ride-hailing services such as Uber have becoming an important part of the urban mobility landscape in many cities around the world. Their operation has been accompanied by fervent debate. In regard to its interaction with public transport (PT), proponents claim ride-hailing is among the important facilitators of a car-independent lifestyle and hence benefits PT, while opponents argue that ride-hailing primarily cannibalizes PT services.

Previous studies analysed ride-hailing user characteristics from household travel surveys or designated surveys distributed among service users (e.g. Young et al. 2019, Young et al. 2020, Gehrke et al. 2019, Tirachini and del Rio 2019). These studies concur that wealthy travellers aged 20-39 make up the predominant user group of ride-hailing services. In contrast, their findings differ in regard to the impact of ride-hailing on PT trips. Specific trip attributes of the ride-hailing option as well as the mobility alternatives are expected to be more relevant for explaining the potential substitution of ride-hailing than socio-demographic characteristics of the individual (Gehrke et al. 2019). An analysis of Uber data from 24 US regions concluded that areas with diverse land-uses, dense population, employment and PT services are characterized by high Uber demand as well as high overall travel demand (Sabouri et al. 2020). In contrast, Uber demand is, all else being the same, lower in areas having better accessibility by private car and PT, as those are associated with a higher competitiveness of alternative travel modes. PT, especially urban rail, was positively associated with adopting ride-hailing services in large metropolitan areas in the US (Sikder 2019 and Grahn et al. 2019). This illuminates the need to examine the impact of ride-hailing on accessibility in different urban contexts which are characterized by urban rail development levels.

Past studies reached contradictory conclusions concerning the impact of Uber entry to the stagnation or decline in PT ridership reported in major North American cities in recent years (Hall et al. 2018, Boisjoly et al. 2018, Graehler et al. 2019). A Toronto survey revealed that 31% of sampled ride-hailing trips are not more than 15 minutes shorter than their fastest PT alternative, while 27% are more than 30 minutes shorter (Young et al. 2020). For 66% of 283 ride-hailing observations in San Francisco their PT counterparts were at least twice as long, suggesting that PT would not have been a competitive alternative for the majority of the trips (Rayle et al. 2016). Survey data did not contain trip-specific waiting times for ride-hailing trips and hence assumed fixed waiting time, disguising potential disparities in ride-hailing service availability. Moreover, travel costs are estimated rather than observed. The representativeness of the selected samples in the abovementioned surveys is also unknown.

Travel times by PT versus car in four cities have been compared based on traffic data and PT timetables while travel demand has been estimated based on Twitter data (Liao et al. 2020). They found that travelling by PT takes 1.4-2.6 times longer than driving a car and follows a similar pattern across cities, with the PT alternative being superior in only a fraction of cases. The spatial properties of Uber service provision has been analysed for the cases of Atlanta with findings indicating that Uber's accessibility is positively correlated with population and road network density (Wang and Mu 2018).

The debate has mostly lacked empirical evidence on ride-hailing service attributes and the corresponding PT demand. Consequently, the interaction between the two services and the extent to which ride-hailing trips compete with or complement PT remains largely unknown. Moreover, it is evident from the literature that while empirical studies in the US are still limited and sparse, European evidence is missing.

To this end, our study addresses the following research questions: (i) How does ride-hailing travel time and cost compare to the fastest PT alternative? (ii) What proportion of ride-hailing trips that do not have a viable PT alternative? (iii) How does ride-hailing change overall service accessibility? For all of these, we are also interested in how answers vary between cities as well as between different areas within cities. We do so by leveraging unique empirical data from Uber and PT for six cities: three US cities – Houston, New York City and Washington DC - and three European cities – Amsterdam, Stockholm and Warsaw. We systematically quantify the increased accessibility associated with ride-hailing services in relation to the accessibility offered solely by PT. By fusing demand data for ride-hailing and PT we also address the following question: (iv) what is the relation between demand share and relative competition between the two alternatives?

Method and Data

We conduct our analysis using a unique dataset of a sample of 3.5 million completed trips made with Uber in October 2018 in New York City, Washington DC and Houston in the US and in Amsterdam (the Netherlands), Stockholm (Sweden) and Warsaw (Poland) in Europe. Each trip $i$ is recorded with its respective origin $o_i$ and destination $d_i$ as well as the following set of time stamps: trip request time, trip waiting time denoted by $t_{w,i}^{RH}$ and trip in-vehicle time denoted by $t_{r,i}^{RH}$. Finally, the fare $c_i^{RH}$ is also registered per trip.

For each ride-hailing trip we query Open Trip Planner (http://www.opentripplanner.org) to find its hypothetical PT counterpart. We query using the ride-hailing trip origin, destination and request time by searching through a detailed graph built from the actual PT schedule and detailed walkable network. We allow for an unlimited number of transfers and limit the maximum total walking distance to 2 kilometers and query for the fastest path, i.e. earliest arrival for a departure later than the ride-hailing request time.

For each ride-hailing trip we have therefore a corresponding PT itinerary for which we extract the following information: walk time $t_{a,i}^{PT}$, waiting time $t_{w,i}^{PT}$, in-vehicle time $t_{r,i}^{PT}$, number of transfers $n_i^{PT}$ and trip fare $c_i^{PT}$.

The duration of the different journey components for each Uber trip and corresponding PT trip (including the required number of transfers for PT journeys) is directly available from the Uber dataset and the PT trip planner, respectively. For privacy reasons the coordinates of the origins and destinations are rounded to three digits (which translates to ca. 50m accuracy for the analysed cities) and request time is rounded to 5 minutes intervals (yet the travel and waiting times remain accurate).

Table 1 provides an overview of the data filtering process applied to the dataset provided. Trips with either an origin or a destination outside of the geo-fenced area (urban PT service coverage area) of each city are removed, as well as trips with missing data. In addition, outlier trips are identified in terms of PT trip duration, Uber fare (average km-fare), Uber trip speed, and the ratio between PT and Uber trip duration. Trips with values larger than three times the standard deviation from the mean for any of these variables were removed. At last, Uber trips for which no PT trip could be found based on the search criteria applied to the PT trip planner were removed. Whilst the latter criterion ranges between 0.2-0.5% for Amsterdam and Stockholm and 1.3% for Warsaw and Washington, this percentage is notably higher for Houston (5.8%) and New York City (6.9%), thus suggesting Uber usage in areas or during time periods (e.g. in nights) where there is no PT service available within a 2 kilometres walking distance.

**Table 1: Summary of data filtering statistics**

|  | Total | Amsterdam | Houston | New York City | Stockholm | Warsaw | Washington DC |
|---|---|---|---|---|---|---|---|
| Initial Uber trips in dataset | 3,506,592 | 113,066 | 288,154 | 2,193,552 | 58,787 | 170,048 | 682,985 |
| Trips outside city geo-fence | 0.0% | 0.0% | 0.2% | 0.0% | 0.0% | 0.1% | 0.0% |
| Missing data | 0.5% | 0.0% | 0.1% | 0.2% | 0.0% | 0.0% | 2.1% |
| Outlier trips | 2.9% | 2.7% | 2.7% | 3.1% | 2.4% | 2.3% | 2.8% |
| Uber trip with no PT match | 5.1% | 0.5% | 5.8% | 6.9% | 0.2% | 1.3% | 1.3% |
| Total trip filtering | 8.6% | 3.2% | 8.6% | 10.1% | 2.7% | 3.7% | 6.2% |
| Final trips in dataset | 3,204,262 | 109,426 | 262,721 | 1,970,968 | 57,191 | 163,592 | 640,364 |

We calculate three journey metrics for each ride-hailing trip and its PT counterpart: nominal travel time $T_i$, generalized travel time $G_i$ and generalized travel cost $C_i$. While the nominal travel time $T_i$ indicates the actual time spent, it is further generalized with relative discomfort ($\beta$'s) attached to the respective trip stages in $G_i$ and then includes the trip fare by translating it into time unit using the value-of-time $\beta_{VoT}$ in $C_i$ (similarly to El-Geneidy et al. 2016). Each of these indicators is computed for the actual trip made with the ride-hailing service as well as for its hypothetical PT counterpart. For ride-hailing the nominal and generalized travel times are composed of waiting and travel time, while for PT it also includes walk time and (for generalized travel time) the number of transfers. For ride-hailing trips these metrics are calculated as follows:

$$T_i^{RH} = t_{w,i}^{RH} + t_{r,i}^{RH} \tag{1}$$
$$G_i^{RH} = \beta_w \, t_{w,i}^{RH} + \beta_r \, t_{r,i}^{RH} \tag{2}$$
$$C_i^{RH} = G_i^{RH} + c_i^{RH} / \beta_{VoT} \tag{3}$$

And similarly for PT:

$$T_i^{PT} = t_{a,i}^{PT} + t_{w,i}^{PT} + t_{r,i}^{PT} \tag{4}$$
$$G_i^{PT} = \beta_a \, t_{a,i}^{PT} + \beta_w \, t_{w,i}^{PT} + \beta_r \, t_{r,i}^{PT} + \beta_n \, n_i^{PT} \tag{5}$$
$$C_i^{PT} = G_i^{PT} + c_i^{PT} / \beta_{VoT} \tag{6}$$

The perceived travel time coefficients ($\beta$'s) are adopted from observed route choices available from smart card data (Yap et al. 2020) as follows: $\beta_a = \beta_w = 1.5$, $\beta_r = 1$, $\beta_n = 5.2$. There is lack of empirical knowledge concerning the perceptions of ride-hailing travel components. We therefore resort to the conservative assumption that the waiting time coefficients for Uber and PT trips are equal.

To compute generalized travel costs, $C_i$, the trip fees are divided by the average Value-of-Time values of the respective country. VoT values in Table 2 are expressed in the local currency (USD, EUR, SEK, PLN). The fare of each Uber trip is readily available in the dataset provided. For PT trips, the fare system of each particular city is considered to estimate the corresponding fare. Depending on the city, a flat fare (Houston, New York City, Stockholm), distance-based fare (Amsterdam), zonal fare (Warsaw) or a combination of flat and distance-based fare (Washington DC) is applied. Fare caps are included where applicable. All fares in place as per October 2018 are used, to provide an adequate comparison with the corresponding Uber trip fare (Table 2). We apply the fare calculation using the fee for a single PT trip. This conservative assumption implies that PT fares for regular users are overestimated, as reduced trip fares resulting from monthly or annual passes are not taken into consideration. Furthermore, we do not consider concessions for different age groups (e.g. the elderly or children), as no age information of individual Uber riders is available. However, age based fare discounts can be expected to typically not apply for the average ride-hailing user (e.g. Sikder 2019).

**Table 2: Summary of case-specific settings**

|  | Amsterdam | Houston | New York City | Stockholm | Warsaw | Washington DC |
|---|---|---|---|---|---|---|
| Value-of-Time | 10.63 EUR | 14.67 USD | 14.67 USD | 125.63 SEK | 45.21 PLN | 14.67 USD |
| PT fare scheme | distance | flat | flat[1] | flat | zonal | distance (subway) flat (bus)[2] |
| PT fare calculation | €0.90 (base) + €0.155*km (km-fare) | $1.25 | $2.75 | 32 SEK | 3.40 zl (20 min, zone 1+2) 4.40 zl (zone 1) 7.00 zl (zone 1+2) | Subway peak: $2.25+$0.675*km [3] Subway off-peak: $2.00+$0.348*km [3] Bus: $2 per trip Transfer subway-metro: $0.50 discount |

[1] A maximum of 1 bus-bus transfer, or 1 bus-subway / subway-bus transfer is allowed per ticket
[2] Fares are capped at $6 (peak, subway only), $3.85 (off-peak, subway only), $7.50 (peak, subway+bus), $5.35 (off-peak, subway+bus)
[3] The (off-) peak fare per km applied for Washington DC subway is an average, as exact fares differ between each station pair in the subway network

Now that we have the abovementioned metrics for the two modes, we can compare them to assess their relative competitiveness. Note that the fare comparison assumes that a single traveller bears the cost. There is a wide-range of measures used in literature to quantify accessibility. We adopt the Modal Accessibility Gap (MAG), an index as proposed by Kwok et al. (Kwok and Yeh 2004) which reflects the difference in accessibility offered by a pair of modes, normalized and bounded by [-1,+1]. We apply this transformation for each of the three metrics as follows:

$$\Delta X_i = (X_i^{RH} - X_i^{PT})/(X_i^{RH} + X_i^{PT}) \tag{7}$$

where $X$ can be substituted by $T$, $G$ or $C$ to obtain the nominal travel time, generalized travel time or generalized travel cost MAG values, respectively. The values of $\Delta X_i$ are null for trips of equal costs, approach -1 for trips where ride-hailing is far more competitive than PT and approach 1 for trips where PT is far more competitive than ride-hailing.

Finally, to quantify the added-value of ride-hailing to the service accessibility offered by PT, we examine how including a ride-hailing alternative improves the (dis)utility (expressed with exponent of nominal travel time $T$ weighted with $\beta$ = -1). We calculate the total utility of two modes and compare it with utility of PT only:

$$A_i^{RH} = ln(exp\,(\beta T_i^{RH}) + exp\,(\beta T_i^{PT})) - ln(exp\,(\beta T_i^{PT})) \tag{8}$$

where utilities (exponents) are first summed and then logarithmized for comparison. Eq. 8 yields positive values growing with added-value of competitive RH trips and null for cases where the RH alternative is not competitive with PT. For illustration, if a certain PT trip takes 30 minutes and the counterpart ride-hailing trip is 15 minutes shorter (i.e. takes 15 minutes), then the increased accessibility is assessed as $A_i^{RH} = 15$. In contrast, if the ride-hailing trip takes 15 minutes longer (i.e. 45 minutes) then there is no improvement in accessibility (i.e. $A_i^{RH} = 0$). If the ride-hailing alternative offers the same travel time (i.e. 30 minutes) then it is considered to add marginally to service accessibility: $A_i^{RH} = 0.7$.

To synthesize and visualize our results both Uber and PT trips are clustered based on their origin location using a hexagonal hierarchical clustering approach H3 (Uber 2018). Each trip origin is assigned to a spatial hexagon of a given aperture. For the set of trips originating from each hexagon, we compute mean values of all three journey metrics, the respective MAG values and the resulting increased service accessibility. To perform cross-city comparison we used aperture of level 9 (hexagons of ca. 300m size), and further refined for the NYC dataset into level of 10. Larger hexagons of level 7 are used for the analysis of demand data in DC.

We perform an additional analysis for the case of Washington DC based on smart card data for the same study period. The data is aggregated to monthly totals to analyse general patterns in terms of mode distribution and demand share between ride-hailing and conventional PT trips. PT data contains both subway and bus ridership data for WMATA services, which are linked to total journeys in the DC area using the Card ID of the passenger and a transfer inference algorithm to determine whether two separate trips should be linked to one integrated PT journey (for details see Krishnakumari et al. 2020). As the ride-hailing data available for this study is a representative sample of the total ride-hailing dataset, the sample ride-hailing trips are uplifted to 100%. Both PT and ride-hailing trips are clustered based on their origin location using a hexagonal hierarchical clustering approach H3 (Uber 2018). By adopting a lower spatial resolution, we make sure that the ride-hailing sample is also representative for the different spatial clusters.

**Results and Analysis**

To compare Uber and PT services, we first describe them in the six analysed cities by examining the distributions of five introduced trip metrics. We use travel time comparisons to estimate the proportion of ride-hailing trips with no viable PT alternative. Next, we visualize modal competition using the modal accessibility gap (MAG) index and its distributions. We then use accessibility maps to visualize the spatial variations in modal competition and the added-value of ride-hailing. Finally, using additional PT demand data for Washington DC, we analyse how the relative competition between ride-hailing and PT impacts the spatial differences observed in their respective demand shares.

*Comparison of Public transport and Uber journey metrics*

We first compare travel times between ride-hailing and the fastest PT alternative for each trip. To this end, we analyse the distribution of four distinct travel time metrics: out-of-vehicle, in-vehicle, nominal and generalized travel time. Out-of-vehicle (i.e. walking time and waiting time in the case of PT and waiting time for Uber services) and in-vehicle travel times are underlying components of nominal travel time $T_i$ which is the actual time spent on a trip. Nominal travel time is generalized by weighting each travel component with its relative discomfort to get the generalized travel time $G_i$. For example, waiting times are perceived more negatively than in-vehicle travel times. We then calculate generalized travel cost $C_i$ of the two modes by incorporating trip fare in the generalized travel time.

The results are presented in Figure 1. In general, it is evident that travel times, and in particular the out-of-vehicle time components, are shorter for the observed Uber trips (trips o Uber's platform) than those associated with the fastest PT alternative. The proportion of travel time that is out-of-vehicle is also lower for ride-hailing - 23% (Washington DC) to 36% (Amsterdam) - as compared to PT with 44% (Warsaw) to 60% (New York City).

Uber out-of-vehicle times are consistently short in all cities. For example, no more than 2% of the rides experience an out-of-vehicle time longer than 15 minutes. For comparison, the corresponding out-of-vehicle share for PT is between 42% in Amsterdam and 65% in New York City. Out-of-vehicle times in PT are not only typically longer but are also characterized by a longer tail of the distribution, including out-of-vehicle times in excess of 40 minutes.

PT alternatives for Uber trips completed in Stockholm, Amsterdam and New York City offer shorter in-vehicle times than those in Warsaw, Houston and Washington (Figure 1(b)). Interestingly, New York City is the only city amongst those considered in this study for which on average PT in-vehicle times are shorter than those observed for the Uber trips. This can be arguably attributed to a combination of high subway service coverage and roadway congestion.

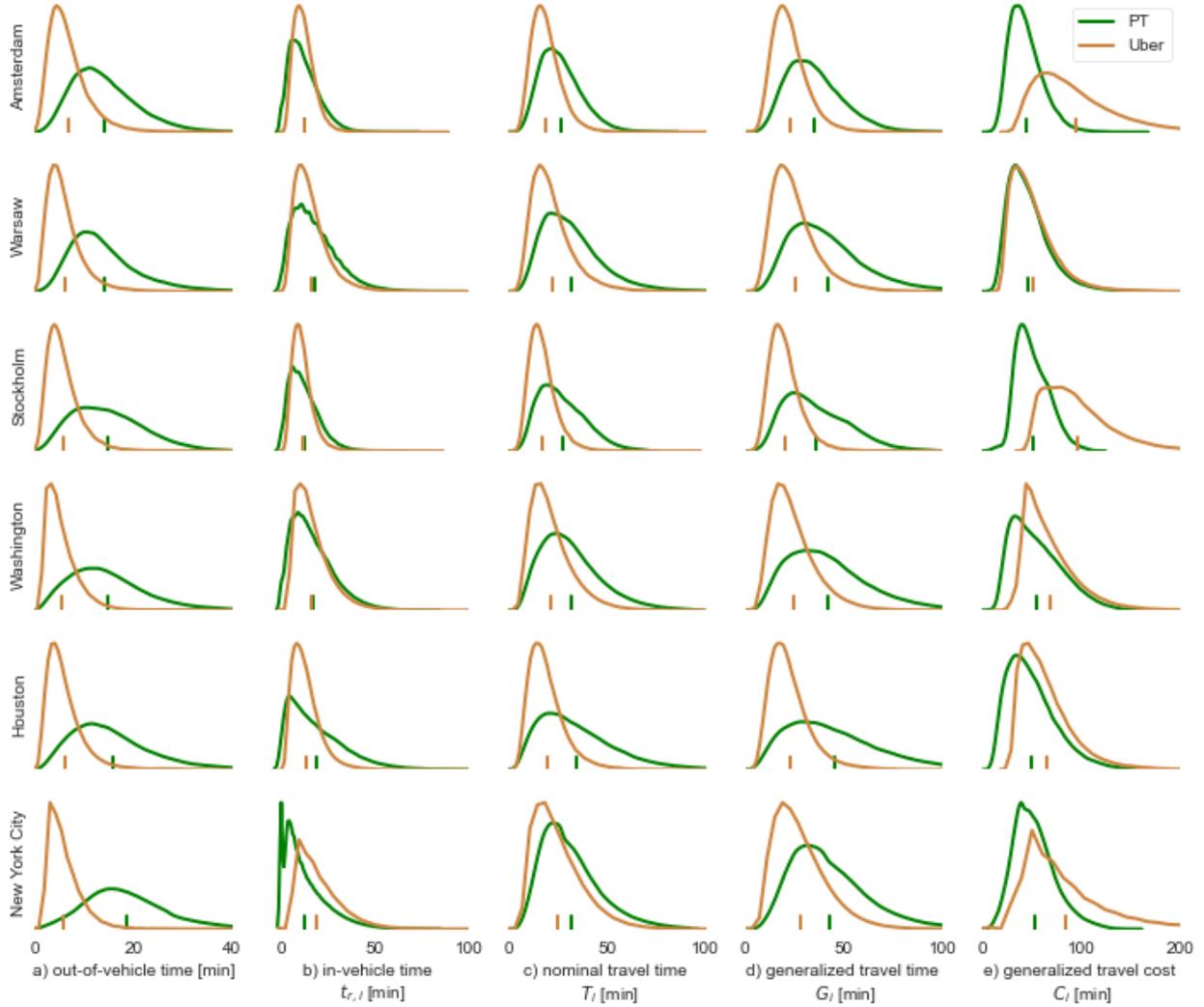

**Figure 1.** Histogram of out-of-vehicle time (a), in-vehicle time (b), nominal journey time (c), generalized travel time (d) and generalized travel cost (e) by public transport and Uber for each of the six cities. Median values are denoted by the x-axis ticks.

The combination of the trends observed for out-of-vehicle time and in-vehicle time is manifested in the nominal travel time distributions (Figure 1(c)). The total travel time is generally shorter for Uber rides. Between 65% (New York City) and 84% (Amsterdam) of Uber rides in all cities are between 10 and 30 minutes. The corresponding range for PT trips is between 44% (Houston) and 65% (Amsterdam). Houston has the longest tail of PT travel times whereas Amsterdam has the narrowest distribution, arguably due to the higher urban sprawl of the former and better PT in the latter. The generalized travel time distributions (Figure 1(d)) overall follow a very similar pattern. The main difference being that PT values become more dispersed since out-of-vehicle times are assigned with a higher weight. For brevity, we will not include the generalized travel time in subsequent analysis (note that it is included in the generalized travel cost metric, Eq.3 and Eq. 6).

While trips on Uber often have shorter waiting and in-vehicle times along with hardly any walking compared to the fastest available PT alternative, it comes at a cost, i.e. a higher price. This is clearly visible in our analysis of the generalized travel cost (Figure 1(e)) which incorporates the travel fee translated into time units based on the local value-of-time. The combination of short generalized travel time and low ticket fees for PT results in low generalized travel costs for Amsterdam and Warsaw, whereas the opposite is observed for Stockholm and Washington. The high Uber fares in Stockholm and Amsterdam more than counteract their short generalized travel time, resulting with high generalized travel costs. Only in Warsaw, where Uber fares are relatively low, the generalized travel cost distribution for Uber is on par with the one obtained for the counterpart PT rides.

*Modal Accessibility Gap analysis*
We are first interested in identifying what proportion of ride-hailing trips have a viable PT alternative, our second research question. We assess the overall level of competitiveness of PT in a city compared to ride-hailing by identifying the prevalence of cases where a ride on Uber would have been excessively long if it was to be performed by PT. We compute thus the ratio between the nominal journey time for PT and Uber (i.e. $T_i^{PT}/T_i^{RH}$) for each of the trips in our dataset and plot the cumulative distribution functions in Figure 2(a). Dutch research suggests that when this ratio exceeds a value of 2.0 - i.e. indicating that the PT trip is at least twice as long as the corresponding car or ride-hailing trip - PT is generally considered an unattractive alternative for non-captive users (Rijkswaterstaat 1995). Consequently, no modal shift is typically observed for ratios reflecting more than a doubling of the travel time. For our six case study cities, the percentage of trips where the ratio exceeds 2.0 is 13.4% (New York City), 16.1% (Amsterdam), 18.9% (Washington DC), 19.1% (Warsaw), 23.6% (Stockholm) and 35.9% (Houston) (see Figure 2c). Therefore, between one out of eight (New York City) and one of three (Houston) ride-hailing trips have no available attractive corresponding PT alternative, for example due to long waiting times during off peak periods. These percentages can be added to those that could not be matched with a viable PT alternative within acceptable walking distances (see Table 1 in the Methods and Data section). The rest of the ride-hailing trips - a vast majority - have a viable PT alternative within acceptable walking distance. It is expected that areas where these trips occurred are those where ride-hailing contributes most to improved service accessibility as discussed in the following sub-section. Overall, similar patterns are obtained for all six cities. In all cities, a sizable minority of the trips – ranging from 15% for New York City and Amsterdam to 8.7% for Houston - has a PT alternative that is time-wise more attractive than the Uber counterpart.

Next, we define the MAG index by contrasting the two modes – PT and ride-hailing – quantifying thus the difference in accessibility offered by the two modes. The index ranges between -1 when ride-hailing is far more competitive and +1 when PT is far more competitive. A MAG value of zero means that both alternatives are equally competitive (see Method and Data). In line with the results reported above, the ride-hailing alternative is found more attractive (MAG<0) for the vast majority of all trips when solely comparing travel times, see Figure 2(b) which presents the distribution of $\Delta T_i$, where each observation corresponds to the travel time of an Uber trip and the corresponding PT trip.

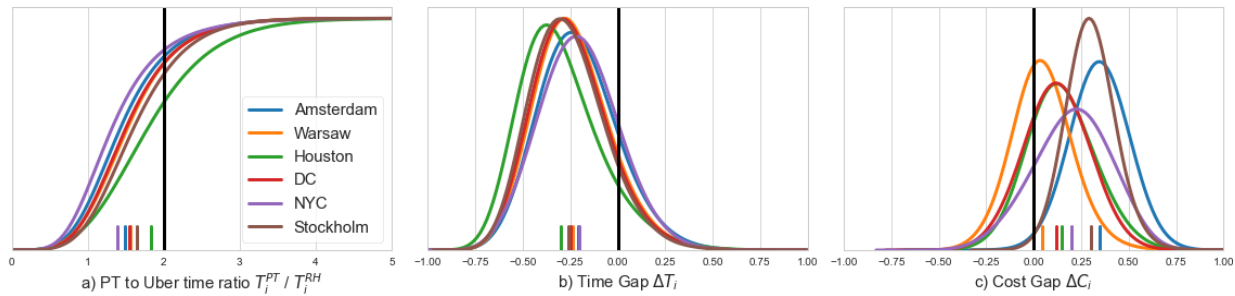

**Figure 2. The cumulative density function of the ratio between the nominal journey time of public transport and ride-hailing (a), and the histograms of the Modal Accessibility Gap (positive for competitive PT and negative otherwise) between public transport and ride-hailing in terms of (b) nominal journey time and (c) generalized travel cost**

The position of PT and Uber changes dramatically once trip fares are accounted for. Moreover, the consideration of fare varies significantly for different cities. Figure 2(c) presents the distribution of $\Delta C_i$, generalized travel costs under the assumption that a single traveller bears the cost. For all cities, the distribution shifts to the right with most observations having a positive gap value. This means that in all cities the majority of the observed Uber trips have a PT alternative which is more attractive than the Uber ride made when considering both time and cost aspects. Riders nevertheless opted for Uber. This presumably stems from individual preferences and attributes such as the importance of comfort or a higher than average value-of-time due to trip purpose, scheduling constraints or disposable income. Urban environment or weather conditions may also cause a higher penalty for walking, waiting and transfers than assumed here based on average estimates. This might also be partially explained by the difference in how PT and ride-hailing fares scale, i.e. the latter might be divided among several travellers in some cases.

In contrast to the nominal travel time MAG in Figure 2(b), pronounced differences amongst cities can be observed for the generalized travel cost MAG values presented in Figure 2(c). The relatively cheap Uber trips in Warsaw (under 5 times more than PT) results with Uber being overall a more competitive alternative for a large minority (40%) of the cases. In contrast, Uber fares in Amsterdam and Houston are significantly higher (more than 8 times) than their PT counterparts, resulting in large gaps between PT and ride-hailing for the majority of trips. When comparing the histograms in Figure 2(b) and 2(c) it is evident they differ not only in their central and range of values but also in their spread, with the generalized travel cost gap function being considerably narrower than the nominal time accessibility gap. This stems from trip fares having smaller variability than travel times as both Uber and PT fees include fixed elements or in the case of the PT services are sometimes strictly fixed (see Method and Data section).

*Spatial distributions of accessibility measures*
We now turn to analysing the spatial patterns of PT and Uber accessibility measures and the resulting impact of ride-hailing on overall service accessibility. To this end, we overlay a hexagonal grid of ca. 200m per side using H3, Uber's Hexagonal Hierarchical Spatial Index. The value of each cell corresponds to the average value for all Uber trips originating from this cell. Note that as the following grid cells in all plots in Figure 3 and across cities are of equal hexagons aperture size (level 9), city maps are directly comparable in terms of their spatial dimensions. The trip data from NYC allows for a spatial analysis at a greater level of detail (hexagons aperture level 10) as presented in Figure 4, providing a finer resolution of columns (d) and (e) in Figure 3.

We first plot the average travel time (i.e. nominal journey time) for PT and Uber, see columns (a) and (b) in Figure 3. Uber's service provides high accessibility (i.e. short travel times) almost ubiquitously. This extends beyond the areas where PT accessibility is at its best, for e.g. city centers and other areas with high activity density. PT accessibility is to a large extent determined by the underlying urban rail network (also shown in Figures 3 and 4) since those offer short waiting times and fast connections. This is particularly visible in the cases of Stockholm, Washington and New York City where corridors of lower travel times can be observed.

Next, we examine the gap between PT and Uber services in terms of nominal journey time and generalized travel cost, aggregated to grid cells. The spatial distribution of these gaps are displayed in columns (c) and (d) of Figure 3, respectively. The striking differences observed between the MAG values for nominal journey time and generalized travel costs in Figures 2(a) and 2(b), respectively, are clearly echoed when comparing Figures 3(c) and 3(d). Note that both MAG plots are displayed using the same colour scheme. Average travel time by PT is longer for all origin cells in all six cities without any exception (Figure 3(c)). The average time gap value per trip ranges from -0.3 for Houston to -0.2 for New York City. Gaps are however small for most origins in Amsterdam and Warsaw as well as for large parts of Washington and New York City, whereas in Stockholm those are limited to part of the inner-city and such zones barely exist in Houston. The spatial distributions thus reveal distinctive patterns amongst the case study cities even though differences were hardly visible when analysing the overall distributions (Figure 2(a)).

A completely different picture emerges when MAG is analysed in terms of generalized travel cost (Figure 3(d)). Average generalized travel cost by Uber is worse off than its PT counterpart for almost all origin cells in all cities, with few notable exceptions in the periphery of Houston and Washington, and to a lesser extent also in Warsaw and New York City. Particularly high gaps in favour of PT are observed in the vast majority of the Amsterdam case study, the inner-city of Stockholm, and in Manhattan for New York City.

The final column (e) in Figure 3 depicts the increase in service accessibility attributed to Uber trips in relation to the PT service provision available in each origin zone for each of the networks based on the log-sum formulation (Eq. 8) using the nominal journey times terms for each mode (Figure 3(a) and Figure 3(b)). A value of zero implies that Uber on average does not offer any travel time benefits while larger values indicate increasing improvements in travel accessibility (see definition in the *Method and Data* section). It is evident that the added-value of Uber to service accessibility varies greatly between cities as well as within cities. Uber delivers most increase in service accessibility in areas which are not well served by urban rail. In our US case study cities, this includes the entirety of Houston with the exception of the downtown area, Southeast Washington and Virginia, and in New York City the Bronx and parts of Queens. This can be clearly seen in Figure 4 (right) where areas which are peripheral and/or with low rail network density are those that see a substantial increase in accessibility due to Uber. In contrast, in our case study European cities such areas are confined to the furthermost areas of Amsterdam (e.g. the harbour), south-western and northernmost suburbs of Warsaw and areas that fall between metro and commuter train corridors in the south-eastern and north-eastern districts of Stockholm. Moreover, Uber also offers an increased accessibility for some sub-centre zones which are located in proximity to urban rail stations but where these services are limited to radial connections. The added-value to service accessibility stems from the competitive travel times for non-radial trips originating from these areas for which PT will often involve longer waiting times, detours and transfers.

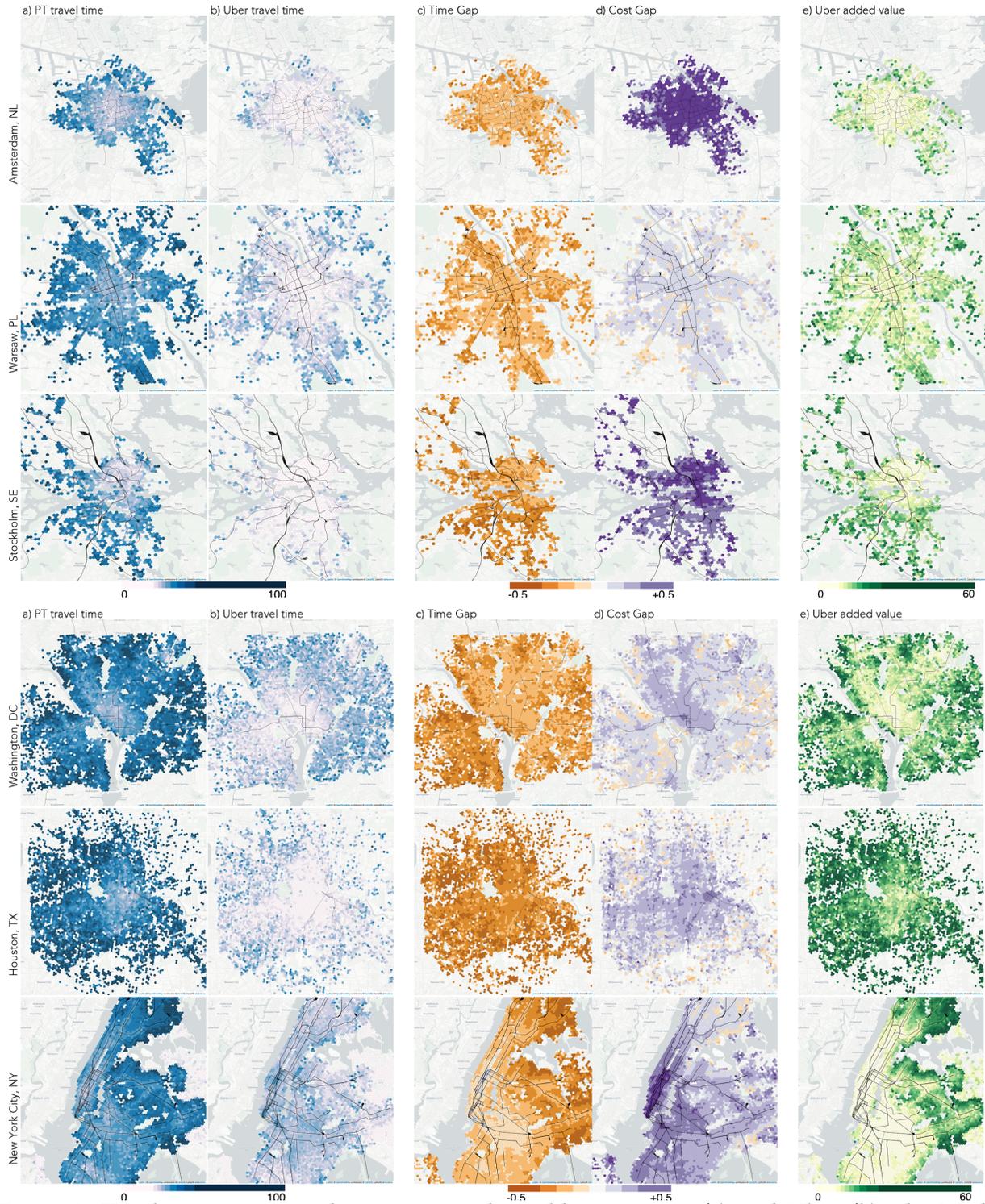

Figure 3. Zonal average nominal journey time by public transport (a) and Uber (b); the Modal Accessibility Gap between public transport and ride-hailing for the nominal journey time (c) and generalized travel cost (d); and the increase in service accessibility associated with Uber (e) for Amsterdam, Warsaw and Stockholm (above), and Washington, Houston and New York City (below), urban rail corridors (tram, light rail, subway, commuter train) are marked in black.

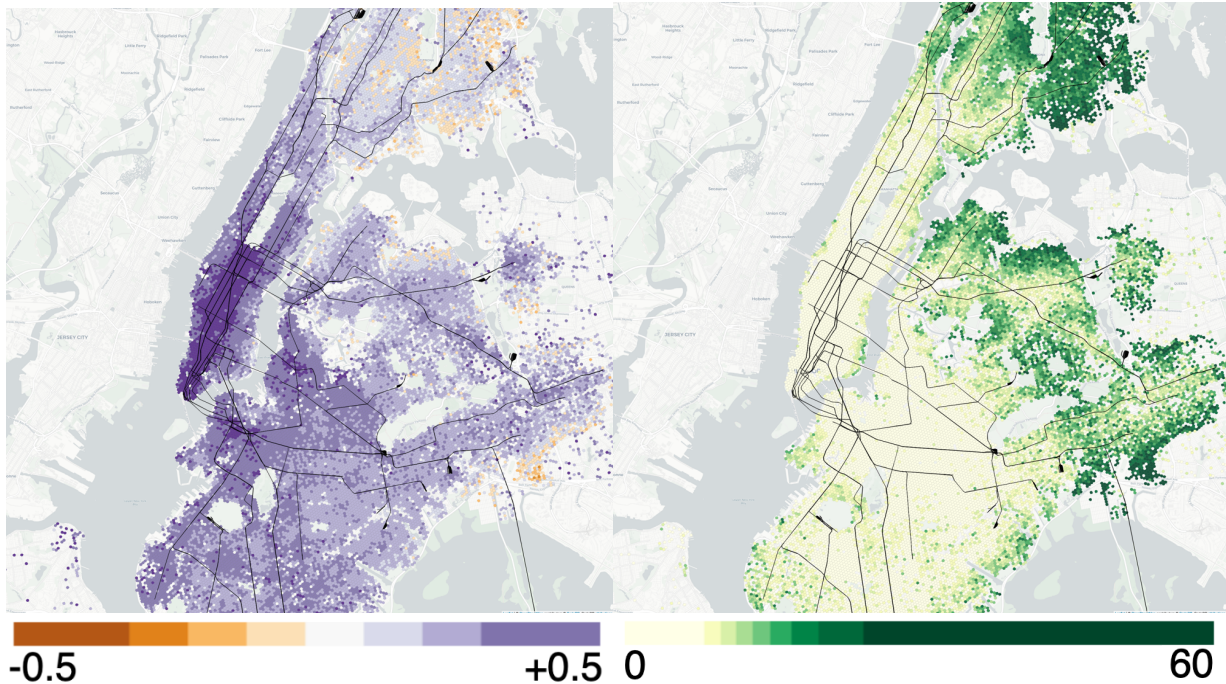

**Figure 4.** Zonal Modal Accessibility Gap between public transport and ride-hailing in New York City for generalized travel cost (left) and increase in service accessibility associated with Uber (right) with a high spatial resolution

*Public transport and Uber demand shares*

We expect the spatial variations in service accessibility for Uber versus PT to result in different travel demand shares as a result of relative attractiveness. We illustrate this by analysing the demand split between PT trips offered by the Washington Metropolitan Area Transit Authority (WMATA) and ride-hailing Uber trips, thereby addressing our fourth research question. We perform our analysis for one of the case study cities, Washington DC, for which we have access to individual smart card transactions (for the description of the data and the selection of analysis resolution see the *Methods and Data* section). For this purpose, we leverage on automated fare collection (AFC) data and fuse it with the ride-hailing data (see Methods section). The modal share of Uber in relation to WMATA is shown in Figure 5. Note that only trips performed by these two operators are considered here (i.e. excluding trips performed by other (ride-hailing) service providers, taxi, private car or walking), whose combined market share in the Washington DC area is estimated to amount to roughly 20% (Deloitte 108, Statista 2020). These results can be examined against the background of the MAG results presented in Figure 3.

The patterns observed in Figure 5 indicate the demand share of Uber rides varies significantly for trips originating from different parts of the metropolitan area. The share of ride-hailing trips is particularly high in areas which are not in direct proximity to a subway service. Most notably, this applies to the areas in the north-western part of DC on both shores of the Potomac River (where Figure 3 already indicated a relatively large time gap between ride-hailing and its PT alternative), as well as trips originating from the East Potomac Park island (middle of Figure 5), where PT services are very limited. Furthermore, we can observe relatively high ride-hailing demand shares for zones close to the Capital Beltway ring-road surrounding DC (I-495), where the road network provides a more competitive alternative to the PT network. Notwithstanding, exceptions to this rule can be presumably attributed to socio-economic variables such as disposable income and motorisation rates. In contrast, areas with a relatively high PT share are clearly visible around the different subway lines, illustrating the positive impact subway lines have on PT mode share, also visible in the results of the modal accessibility gap analysis in Figure 3.

Overall, this map suggests a rather limited competition between ride-hailing and rail-based conventional PT: the share of ride-hailing trips is mainly high for areas where rail-based PT does not provide a reasonable alternative. Competition between ride-hailing and conventional buses can however not be ruled out based on Figure 5.

We find an overall negative correlation of -0.492 between the MAG of nominal travel time and the respective Uber share at the hex zonal level (see Figure 6). This means that for trips for which PT is competitive, the share of ride-hailing is low and vice-versa. While this confirms the expected relation between accessibility and demand, it also demonstrates that much of the spatial variability in Uber demand share cannot be solely explained by spatial discrepancies in accessibility.

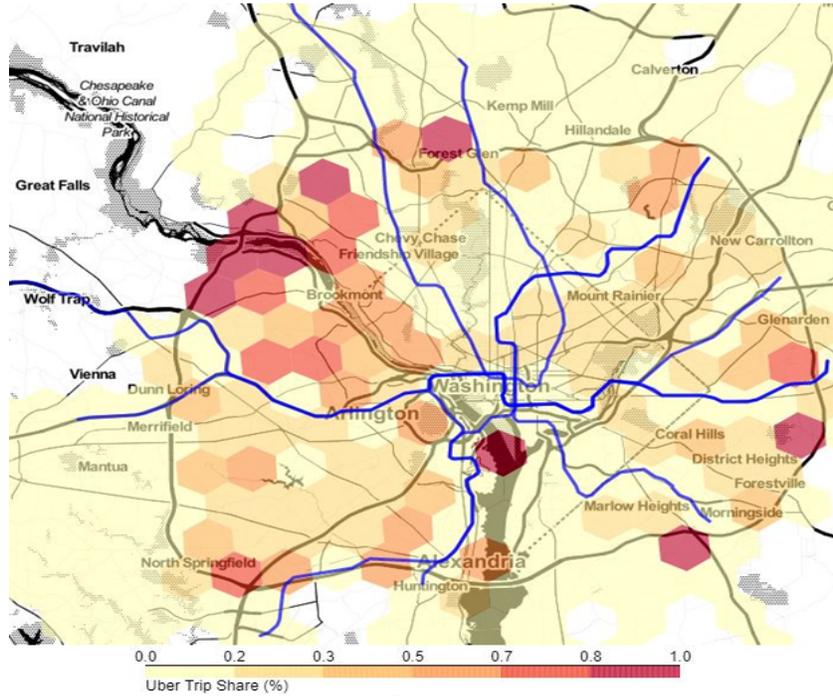

Figure 5. Uber demand share as a percentage of Uber and public transport demand in Washington DC, based on trip origin. Blue lines reflect the different metro lines, and darker red colours indicate a higher Uber demand share.

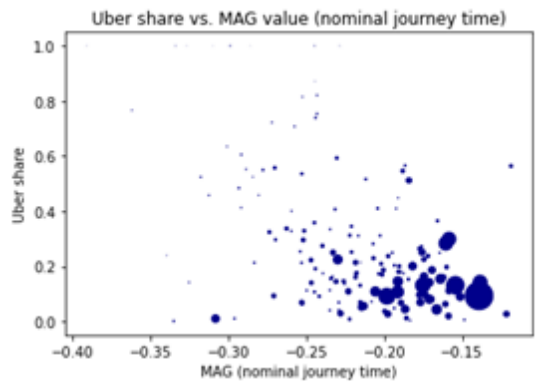

Figure 6: Scatter plots relations between the share of Uber demand (as a fraction of Uber and public transport demand) and the Modal Accessibility Gap for nominal journey time. Each dot represents a hex in Washington DC and its size corresponds to the total demand for public transport and Uber originating from that zone.

Concluding Discussion

Over the last decade, ride-hailing has become an important travel alternative in many cities worldwide. Planners and policy makers have been interested in better understanding this new mobility option. Of particular interest is it's interaction with existing public transport services and whether these two options compete with or complement each other. In this study we provide an evidence-based account and our findings point to a nuanced reality. We use Uber trip data in six cities in the US and Europe to identify the most attractive public transport alternative for each individual ride. We then use this data to compare the two modes, calculate modal accessibility gap (MAG) indices and study their spatial distribution, and finally assess how relative attractiveness of the two options relates to their demand share.

In response to our first research question, i.e. *how does ride-hailing travel time compare to the fastest public transport alternative?*, our findings show that for most trips across the six cities, shorter out-of-vehicle and in-vehicle travel times are observed on Uber trips than their fastest public transport counterpart. For 13-36% of the Uber trips, the travel time associated with ride-hailing is at least twice as fast. Much of these differences stem from out-of-vehicle time (walking and waiting) which dominates the difference between PT and Uber. Moreover, for 0-7% of the Uber trips no viable public transport alternative within a reasonable walking distance could be found. The combination of these two elements - ranging between 20% and 40% - answers our second research question, i.e. *what proportion of ride-hailing trips do not have a viable public transport alternative?* Notwithstanding, across the six studied cities, for 9-15% of the Uber rides we could actually find a shorter public transport ride.

These findings highlight that Uber services are used in both competing and complementary circumstances. An almost mirrored picture emerges when including travel fees in the modal accessibility gap analysis. While offering cheaper fares than conventional taxis, Uber is still more expensive than public transport and offers time savings at a higher cost. The cost difference is on average larger than the time difference when converting the latter using the local value-of-time. In other words, for the average trip, using the average local value-of-time, the time savings offered by Uber are not worth the added monetary cost. However, all observations in our sample pertain to cases where individuals opted for the ride-hailing alternative. This can be explained by ride-hailing customers having different monetary evaluation of their time, attach value to factors which are not considered in this study such as comfort, safety and reliability or would not have considered public transport as an alternative. These aspects should be addressed in future travel behaviour research focused on ride-hailing and mode choice. Lastly, the attractiveness of ride-hailing will increase substantially if more than one rider is considered in this analysis.

We find that the answer to our third research question - i.e. *what is the impact of ride-hailing on overall service accessibility?*- varies greatly within as well as between cities. In general, we find that the increased service accessibility - as measured in terms of travel times - attributed to the inclusion of ride-hailing is particularly high in areas without urban rail service. This results in an overall greater added-value of accessibility in our US case study cities than in the European cities. There are however also large differences within cities. For example, large parts of New York City (i.e. Manhattan and Brooklyn) having similar (high) MAG values which reflect lesser advantage for Uber and (low) increased service accessibility levels that can be attributed to Uber, both of which are comparable to those observed for most areas of Amsterdam. At the same time, outer Queens exhibits similar levels to those observed for the majority of Houston. Conversely, the outskirts of Warsaw and Stockholm exhibit patterns comparable to those of the cells lying just outside the downtown areas in Washington and Houston. These differences are arguably driven by spatial variation in public transport and Uber services' availability. The former is largely determined by network geometry, which tends to focus on radial connections and has a higher density in the core. In contrast, Uber availability is decentralized with fares depending on labour costs and travel times largely determined by road network speeds and congestion.

We finally examine *what is the relation relative travel time competitiveness and respective demand share.* The overall travel time accessibility gap is echoed in the demand shares from our analysis fusing PT and Uber demand data for Washington DC. The share of ride-hailing is low when public transport offers a competitive travel alternative and vice-versa. Hence, our findings point to a more nuanced reality than the one implied by the dichotomy of the complementing versus competing discourse.

This study contributes to the growing empirical evidence on the interaction of ride-hailing and public transport in cities worldwide. Our findings can support planning and policy decisions, such as identifying gaps in the public transport network and addressing spatial disparities in service provision. Future work may shed light on the importance of temporal and socioeconomic factors in explaining the observed patterns. Differences between private and pooled ride-hailing rides (such as UberPool) may also be a subject of future research and provide insights on the relative positioning of shared rides in relation to private rides and public transport. Our analysis is performed based on part of the travel demand - namely Uber trips, rather than the entire urban travel demand. Further analysis may also examine demand for other alternatives as well as the role of land-use distribution and socio-economic data to shed light on the underlying determinants of the observed differences between public transport and ride-hailing usage as well as the determinants of their demand.